# Developing and Validating the Arabic Version of the Attitudes Toward Large Language Models Scale

*Basad Barajeeh[1], Ala Yankouskaya[2,*], Sameha AlShakhsi[3], Chun Sing Maxwell Ho[1], Guandong Xu[1], Raian Ali[3,*]*

*[1]The Education University of Hong Kong, Tai Po, Hong Kong*
*[2]Faculty of Science and Technology, Bournemouth University, Poole, UK*
*[3]College of Science and Engineering, Hamad Bin Khalifa University, Doha, Qatar*

**Abstract**. As the use of large language models (LLMs) becomes increasingly global, understanding public attitudes toward these systems requires tools that are adapted to local contexts and languages. In the Arab world, LLM adoption has grown rapidly with both globally dominant platforms and regional ones like Fanar and Jais offering Arabic-specific solutions. This highlights the need for culturally and linguistically relevant scales to accurately measure attitudes toward LLMs in the region. Tools assessing attitudes toward artificial intelligence (AI) can provide a base for measuring attitudes specific to LLMs. The 5-item Attitudes Toward Artificial Intelligence (ATAI) scale, which measures two dimensions, the AI Fear and the AI Acceptance, has been recently adopted and adapted to develop new instruments in English using a sample from the UK: the Attitudes Toward General LLMs (AT-GLLM) and Attitudes Toward Primary LLM (AT-PLLM) scales. In this paper, we translate the two scales, AT-GLLM and AT-PLLM, and validate them using a sample of 249 Arabic-speaking adults. The results show that the scale, translated into Arabic, is a reliable and valid tool that can be used for the Arab population and language. Psychometric analyses confirmed a two-factor structure, strong measurement invariance across genders, and good internal reliability (Cronbach's α ranged between .67 and .75). The scales also demonstrated strong convergent and discriminant validity. Our scales will support research in a non-Western context, a much-needed effort to help draw a global picture of LLM perceptions, and will also facilitate localized research and policy-making in the Arab region.

**Corresponding authors:** Correspondence to Raian Ali (raali2@hbku.edu.qa) and Ala Yankouskaya (ayankouskaya@bournemouth.ac.uk)

## 1. Introduction

Research on public attitudes towards artificial intelligence (AI) in Arab populations has expanded in recent years, reflecting the region's rapid digital transformation. Across studies, AI attitudes in this population are generally characterised by a combination of optimism (Liebherr et al., 2025b; Babiker et al., 2025; Rahman et al., 2024). Large-scale survey evidence from multiple Arab countries consistently shows that although reported knowledge of AI tends to be low or moderate, expectations regarding its transformative potential remain high (Al Omari et al., 2024). At the national level, findings from 10 Arab countries indicate relatively widespread awareness of AI: 77% of respondents reported familiarity with the term and high levels of enthusiasm, with eight in ten expressing a positive view of AI-enabled products and services (Syed et al., 2024).

These positive attitudes are further supported by comparative and cross-cultural research. For example, Arab participants have been shown to evaluate AI's impact on personal well-being more favourably than their British counterparts (Liebherr et al., 2025a). Similarly, Naiseh et al. (2025) reported that positive attitudes towards AI were significantly associated with self-efficacy and perceived competency, with Arab respondents expressing stronger expectations of individual benefit. This trend is mirrored in global surveys, which find that populations in emerging economies, including many Arab countries, consistently report higher levels of AI acceptance, trust, and engagement compared to advanced economies in North America and Europe (Gillespie et al., 2025). However, AI is a generic term and can range from simple auto-correction features in an email editor to advanced robotics, as well as finance, stock prediction, and smart traffic solutions. This means that measures for attitudes toward AI should not assume that attitudes toward AI as a monolithic concept will automatically transfer to specific families of AI systems, such as, in the case of this paper, Large Language Models (Liebherr et al., 2025b).

At the same time, attitudes toward AI in the Arab world are marked by ambivalence, as enthusiasm in acceptance is accompanied by fears and anxieties regarding its personal and societal impact. The most prominent concern is job displacement: nearly one-quarter of the Saudi public (24.7%) report fearing job loss due to AI, a concern particularly acute among professionals (Syed et al., 2024; Hassouni & Mellor, 2024). Other apprehensions include the erosion of human relationship and connection (Alshutayli et al., 2024), potential misuse of personal data (Frimpong, 2025), and the belief that over-reliance on AI may undermine human creativity (Khosravi et al., 2024). These concerns suggest that while attitudes toward AI are broadly positive in Arab population, they are also tempered by significant reservations about its long-term implications. It also suggests that factors shaping attitudes in a Western context, such as fear and contributors to it, may apply differently, whether to a lesser or greater extent, in other regions, potentially making some components and questions within attitude measures less applicable.

The ambivalence reported in Arab public attitudes towards AI fits well within established theoretical frameworks. Models like the Technology Acceptance Model (TAM) (Davis, 1989) and its extensions (such as UTAUT) (Davis et al., 1989) suggest that people are more likely to have positive views of a technology if they see it as useful and easy to use, while concerns about risk and trust can lead to fear or hesitation about adopting it (O'Shaughnessy et al., 2023). The Extended Parallel Process Model (EPPM) provides explanation how fear and efficacy appraisals interact: if perceived threats (e.g., job loss, data misuse) are not accompanied by a sense of efficacy (e.g., perceived control over AI use), individuals may resort to fear-control responses such as avoidance or scepticism, even when they recognise potential benefits (Popova, 2012). Similar, the threat-rigidity perspective, which considers this ambivalence within a broader psychological response, argues that individuals may simultaneously recognise both opportunity and threat, resulting in mixed motivations that can reduce their willingness to engage actively with AI (Pei et al., 2025). However, a recent meta-analysis of 243 studies also pointed out that although traditional technology acceptance models still provide a useful base for understanding attitudes towards AI, they fall short when it comes to explaining how people respond to specific types of AI, such as large language models (LLMs) (Chen et al., 2025a). This is because these models tend to focus on practical features and do not fully account for the ways LLMs can trigger emotional responses, affect social interactions, or take on roles like advisor, companion, or personal assistant. For example, users may gradually form parasocial bonds with LLMs due to their capacity to offer emotionally validating responses, sustained availability, and an illusion of reciprocal understanding (Yankouskaya et al., 2025). Therefore, general attitudes towards AI may not apply directly to more complex and socially interactive forms of AI like LLMs.

The extent to which attitudes toward general artificial intelligence overlap with those toward large language models is a complex and context-dependent issue that requires careful empirical investigation. It touches on fundamental challenges in attitude measurement, where the object of study is not only complex but also characterised by fundamental and systematic differences. General AI is often perceived through a lens of societal opinion, framed by media narratives that create a gap between how individuals perceive AI's impact on society at large and how they perceive its impact on their own lives. For instance, recent surveys provide evidence that people are more worried about AI's effect on overall employment than on their own job security (Dreksler et al., 2025). This suggests that public attitudes toward general AI are often based without association with personal circumstances (Naveed, et al., 2023). However, fears that employers and societies may eventually depend on them and replace humans could still apply. Fear components extend beyond the loss of jobs and relate to risks associated with elevated LLM autonomy, such as when they are embedded in critical decision-making systems (Włoch et al., 2025).

In contrast to AI, LLMs are not merely passive instruments or subjects of opinion; they function as highly effective persuasive agents (Yankouskaya et al., 2025). Experimental evidence shows that messages generated by LLMs can produce significant attitude change in humans, comparable to the effect of messages written by people (Bai et al., 2025). Furthermore, people tend to be generally receptive to advice generated by LLMs (Wester et al., 2024). These observations imply that attitudes toward LLMs are more influenced by individual experience and direct interaction than by abstract or second-hand narratives. Direct engagement with LLMs shifts them from an abstract notion to a concrete tool. When people use LLMs to obtain information, generate text, or assist with work, they form judgements based on observable functions and limitations. This interaction gives rise to attitudes that are often more positive and grounded in trust, especially toward the primary LLM used most frequently. As familiarity increases, attitudes tend to become more favourable and more stable. The way LLMs are presented also contributes to this difference. They are frequently described as tools that can increase productivity and support specific tasks (Spotnitz et al., 2024). Such descriptions are understandable and relatable for the general public, who often perceive LLMs as tangible products (Chen et al., 2025b). In contrast, general AI remains a more abstract and less clearly defined concept, often imagined as a single system capable of mastering all tasks. Consequently, attitudes toward LLMs are typically formed in specific usage contexts, whereas attitudes toward general AI tend to be broader and more abstract.

These distinctions suggest that attitudes toward LLMs may not entirely overlap with those toward general AI but could involve a somewhat different evaluative process influenced by users' direct experience, familiarity, and perceived usefulness. This raises an important methodological question: how should such attitudes be measured? A number of established scales have been developed to assess public attitudes toward AI more broadly. For example, the General Attitudes Towards Artificial Intelligence Scale (GAAIS) (Schepman & Rodway, 2020) distinguishes between positive and negative views of AI, including perceived benefits and concerns. The Attitudes Toward Artificial Intelligence Scale (ATAI) offers a more general, bidimensional measure of people's overall stance on AI (Sindermann et al., 2021). The AI Attitude Scale (AIAS-4) includes four concise items assessing perceptions of AI's usefulness, impact, adoption, and risk (Grassini, 2023). Another tool, the ATTARI-12, incorporates cognitive, emotional, and behavioural components to provide a more nuanced profile of AI-related attitudes (Stein et al., 2024). Finally, the Threats of Artificial Intelligence Scale (TAI) focuses on perceived risks across several domains, including employment and health (Kieslich et al., 2021). While the existing instruments offer a foundation for understanding general perceptions of AI, they were not originally designed with LLMs in mind and may overlook important aspects of how people engage with and evaluate these systems. As the use of LLMs becomes increasingly widespread, it is important to examine how attitudes toward them are shaped in different cultural and linguistic contexts. This is especially relevant in Arab populations, where the cultural reception of AI and patterns of technology use may differ significantly from those in Western.

The overarching aim of the present study is to validate two psychometric scales for assessing attitudes toward large language models (LLMs) in an Arab population, by adapting the ATAI framework to the context of LLMs. The original ATAI comprises five items, with three items measuring fear-based concerns about AI and two items assessing acceptance and perceived benefits. Our first proposed scale focuses on attitudes toward LLMs in general (AT-GLLM), capturing broad evaluations of these technologies as a class. The second scale targets attitudes toward the primary LLM that an individual interacts with most frequently (AT-PLLM), recognising that repeated, direct use and familiarity may give rise to more specific and experience-based evaluations. This distinction is theoretically and empirically motivated. General attitudes reflect abstract or societal-level views that may be shaped by media, discourse, and brief user experience with a range of different LLMs (Haensch, 2024; Radivojevic et al., 2024; Wester et al., 2024; Roe & Perkins, 2023; Liu et al., 2025). In contrast, attitudes toward a primary LLM are grounded in frequent interaction with the system, as well as in the perceived usefulness or limitations of that system (Mendel et al., 2025; Enam et al., 2025; Elhassan et al., 2024). Over time, such regular engagement may lead to responses that become almost automatic, as users develop habitual patterns of interaction and form stable impressions. This automaticity may stem from repeated reliance on the system in daily tasks, which reduces the need for active evaluation and reinforces a sense of familiarity and trust.

To ensure the scale remains relevant to the context of large language models, the original ATAI items were adapted by replacing general references to AI with LLM-specific wording (e.g., changing “I trust AI” to “I trust LLMs” or “I trust my primary LLM”). This approach allows the scale to retain the underlying structure and measurement logic of the original instrument, while making the content more specific to the domain of LLMs. In doing so, we aim to balance conceptual consistency with increased relevance to users’ actual experiences. Our study is based on the recently developed and validated attitude scales toward LLMs in a UK population (Liebherr et al., 2025b). This work produced two versions of the scale, targeting general and primary LLMs, and confirmed a two-factor structure consistent with the original ATAI, capturing both acceptance and fear. Based on these findings, we hypothesise that a comparable two-factor solution will emerge in the Arab sample. We further expect that the AT-GLLM scale will show a stronger association with the original ATAI, as both instruments assess generalised attitudes toward AI technologies, independent of specific user experience.

# 2. Method

## 2.1. Study design

The items were adapted from the English version of the AT-GLLM and AT-PLLM scales proposed by Liebherr et al. (2025b) for use in the Arab sample. The development of the scale was part of a larger study aimed at evaluating psychological and behavioural factors related to LLM usage, including attitudes toward LLMs and levels of dependency and relation to personal and use factors. For this study, only the relevant portions of the questionnaire that address the development of the AT-LLM Arabic scale (whether the AT-PLLM or the AT-GLLM) and the variables uses for their external validation, will be used.

At the beginning of the survey, participants were provided with an explanation of large language models (LLMs), including their capabilities and applications. It was emphasized that LLMs extend beyond conversational agents such as ChatGPT and encompass a wide range of functions. The purpose of this introduction was to establish a common baseline of knowledge about LLMs among participants, helping to produce consistent responses related to their familiarity and use.

Following the introduction, participants proceeded to fill out the questionnaires. The initial section of the survey gathered demographic information, including participants’ gender, age, ethnicity, nationality, education level, employment status. To assess actual LLM usage, an additional question asked for a usage frequency rating from 0 to 10. The second section assessed participants’ attitudes toward LLMs, both the primary one and in general, and their attitude toward AI (through ATAI questionnaire). Following this, a single-item questionnaire to assess self-efficacy on a scale from 0 to 10, was asked. The ATAI and self-efficacy scales were incorporated to explore the association between the AT-LLM scales and variables of theoretical relevance.

The AT-LLM scales were translated from their English counterpart proposed in Liebherr et al. (2025b), into Arabic using the back-translation method (Brislin, 1970). First, a bilingual author proficient in both Arabic and English translated the English questionnaire into Arabic. Second, another bilingual author, with expertise in Human-AI interaction and research methodology, reviewed and refined the Arabic translation for accuracy and fluency. Third, a different bilingual academic, unaffiliated with the authorship team and unfamiliar with the AT-LLM scales, independently translated the Arabic version back into English. Finally, the senior author then reviewed and compared the back-translated English version with the original AT-LLM to assess whether the translated version preserved the same meanings as the original.

A pilot study involving 37 participants from the target population was conducted to evaluate the clarity, comprehension, and cultural appropriateness of the translated items and the other questions involved in the larger study. Minor wording adjustments were recommended to better clarify each question's intent and ensure compatibility with Arabic language use. As a result of the translation process and expert validation, the Arabic versions of the AT-LLMs (both the AT-GLLM and AT-PLLM) comprised four subscales and 10 items. Participants rated each item on an 11-point Likert scale, selecting a response from Strongly Disagree (0) to Strongly Agree (10), reflecting their attitude toward each statement. Attitudes comprise two components: Fear and Acceptance. Acceptance assessment is based on two items, while Fear is assessed through three items.

## 2.2. Sample

Participants were recruited via the Prolific online platform (www.prolific.com), which specializes in sourcing participants for research studies, including surveys. The survey was created and distributed using SurveyMonkey (www.surveymonkey.com).

Sample size requirements for validating the Arab sample were derived from the structure and parameter estimates of our UK CFA (12 items, 2 correlated factors, df = 39, loadings .57–.94) (Liebherr et al., 2025b). Using an RMSEA-based approach (MacCallum et al., 1996), we calculated that, for the Arab model (df = 53), detecting a difference between close fit (RMSEA = .05) and not-close fit (RMSEA = .08) at $\alpha = .05$ would require N = 204 for 80% power. Additional calculations showed that, with the smallest UK loading (.57), N = 200 was needed to estimate factor loadings with 95% CI widths $\leq \pm.10$, and N = 220 to estimate $\alpha = .80$ with CI width $\leq .05$ (Bonett, 2002a). We therefore set a target of N = 230, which exceeds the 80% RMSEA power threshold for both models, approaches 90% power for the Arab CFA, and ensures precise estimation of factor loadings, reliability coefficients, and small-to-moderate correlations.

The AT-LLM instrument was initially administered to a sample of 304 Arab-speaking participants. During the cleaning stage, some cases were excluded from the dataset, resulting in a final sample of 249 participants. As all questions were mandatory, there were no missing data to any of the variables of interest to this study. The final sample included 128 participants pursuing or holding postgraduate degrees (51.41%), 97 with undergraduate degrees (38.96%), 15 with vocational or technical education (6.02%), and 9 with secondary education (3.61%).

The average age of participants was 28.31 years (SD = 6.07). 51.8% of participants were female (*n* = 129), while males numbered 119 (47.8%), with 1 participant (0.40%) missing gender data.

## 2.3. Measures

To guide participants, each questionnaire was preceded by a short instruction clarifying its purpose. For the AT-GLLM scale, the participants were told: "*Think of LLMs in general and the applications built on them. To what extent does the following statement apply to you?*" For the AT-PLLM scale, we used the following statement: "*Think of your primary LLM, meaning the LLM you use the most. To what extent does the following statement apply to you?*"

*Attitude Toward General Large Language Models (AT-GLLM)*: This scale measures individuals' attitudes toward LLMs, with items rated on a scale from 0 (strongly disagree) to 10 (strongly agree). It consists of two subscales: acceptance, with two items (e.g., "I trust them"), and fear, with three items (e.g., "I fear them"). Table 1 presents list of items in both English and Arabic.

*Attitude Toward Primary Large Language Models (AT-PLLM)*: This scale consists of five items assessing individuals' attitudes toward primary LLMs. All items use a Likert-type response format ranging from 0 (Strong Disagreement) to 10 (Strong Agreement). The scale measures two aspects of individuals' attitude: acceptance (with two items, e.g., "I trust it") and fear (with three items, e.g., "I fear it").

**Table 1.** Items of the AT-GLLM and AT-PLLM Scales: English and Arabic Versions

| | Scale | Items | Arabic Version |
|---|---|---|---|
| **AT-GLLM** | Attitude toward General LLMs | I fear them | أشعر بالخوف منها |
| | | I trust them | أثق بها |
| | | They will destroy humankind | ستدمّر البشرية |
| | | They will benefit humankind | ستفيد البشرية |
| | | They will cause many job losses | ستتسبب في فقدان العديد من الوظائف |
| **AT-PLLM** | Attitude toward Primary LLM | I fear it | أشعر بالخوف منه |
| | | I trust it | أثق به |
| | | It will destroy humankind | سيدمّر البشرية |
| | | It will benefit humankind | سيفيد البشرية |
| | | it will cause many job losses | سيتسبب في فقدان العديد من الوظائف |

*Attitude Toward Artificial Intelligence (ATAI)*: This scale, developed by Sindermann et al. (2021), assesses participants' attitudes toward AI, with responses on an 11-point scale from 0 (Strongly Disagree) to 10 (Strongly Agree) to each of its five items. It comprises two dimensions: acceptance (two items, e.g., "I trust Artificial Intelligence") and fear (three items, e.g., "I fear Artificial Intelligence"). The subscales demonstrated acceptable reliability, with Cronbach's alpha coefficients of .684 for acceptance and .738 for fear. The AVE values were .522 for acceptance and .537 for fear, indicating adequate convergent validity. The CR values were .680 and .764, respectively. Additionally, McDonald's omega coefficients were .687 for acceptance and .749 for fear, supporting the scale's internal consistency.

*Self-efficacy*: As an external criterion for validation, the GSE-SI (Di et al., 2023), a single-item measure of general self-efficacy, was employed. Participants were asked to rate the statement, "I believe I can succeed at most endeavours I set my mind to," using an 11-point Likert scale from 0 to 10, with higher scores reflecting greater self-efficacy.

## 2.4. Statistical analyses

### *Data preprocessing*

After the initial screening process, the data was cleaned by removing certain cases to ensure its accuracy for analysis. Participants who failed attention check questions were removed from the dataset. Additionally, those who identified as non-binary or chose not to disclose their gender were also excluded due to their small sample size. Furthermore, individuals of non-Arab ethnicity and those who participated multiple times were excluded. Lastly, responses from participants who reported zero frequency of LLM use were omitted, as such responses indicated no experience with or exposure to LLMs, which was necessary for the study's focus.

### *Monotonicity check*

A monotonic relationship between the latent trait and item response is a fundamental assumption in nonparametric item response theory, such as Mokken scale analysis (Sijtsma, & Molenaar, 2002). To evaluate this assumption for AT-GLLM and AT-PLLM scales, nonparametric isotonic regression was utilized, offering a robust and flexible approach that does not require any specific data distribution. The procedure involves calculating the restscore for each item by summing the responses of all other items within the same scale. The restscore is then used as the predictor variable, while the individual item response serves as the outcome variable. By fitting an isotonic regression model, we can determine whether the expected value of an item exhibits a monotonic (increasing or decreasing) relationship with the restscore, thereby assessing the validity of the monotonicity assumption.

For both the AT-GLLM and PLLM scales, the results indicated that the monotonicity assumption held for all items, as no violations were observed. The fitted response scores for the AT-GLLM scale ranged between 0 and 10, with a steady increase across trait range. The item "*I fear them*" exhibited the most significant change, while item "*They will cause job losses*" yielded a small slope. In the AT-PLLM scale, every item displayed a non-decreasing trend, with fitted values between 0 and 10. The strongest relationships with the restscore were observed for "*It will destroy humankind*" and "*It will cause job losses*".

### *Descriptives and correlations*

The first set involved calculating descriptive statistics for each key variable, including means, standard deviations, skewness, and kurtosis. Additionally, Pearson correlation analyses were performed. to explore the relationships between the main variables. The dataset contained no missing responses.

Before conducting the analyses, we assessed the data for deviations from univariate normal distribution. Univariate normality was evaluated using skewness and kurtosis measures, and all variables met the normality criteria, as shown in Table 2 ($|\text{skewness}| \leq 2$; $|\text{kurtosis}| \leq 7$; Hair et al., 2010).

### *Confirmatory factor analysis*

In the second set of analyses, we examined the factor structure of the AT-GLLM and AT-PLLM scales using confirmatory factor analysis (CFA). The primary goal was to verify whether the factor structure identified in previous research (Liebherr et al., 2025b) was replicated in this sample. Model fit was evaluated using several indices, including the Comparative Fit Index (CFI), Tucker–Lewis Index (TLI), Root Mean Square Error of Approximation (RMSEA), and Standardized Root Mean Residual (SRMR). Additionally, the Chi-square test and the Chi-square to degrees of freedom ratio ($\chi^2/df$) were performed. A good fit was indicated by CFI and TLI values of .95 or higher, and RMSEA and SRMR values below .08 (Hu & Bentler, 1999). The maximum likelihood with robust (MLR) was used in CFA as the estimation method. This approach was employed in R software using lavaan package (Rosseel, 2012).

### *Reliability analysis*

The reliability of each construct was evaluated using multiple indices to ensure robustness. Cronbach's alpha was employed to assess internal consistency, with values above 0.7 indicating acceptable reliability. McDonald's omega provided an alternative estimate of internal consistency, offering advantages in cases where item loadings vary, and similarly aimed for values exceeding 0.7. Composite reliability was calculated based on confirmatory factor analysis results, serving as an overall measure of construct reliability, with thresholds above 0.7 denoting good reliability. Additionally, the average variance extracted (AVE) was examined to assess convergent validity, with values of 0.5 or higher indicating that a sufficient proportion of variance in observed variables is explained by the underlying construct.

### *Measurement invariance analysis*

Next, multi-group confirmatory factor analysis (MGCFA) was conducted to assess measurement invariance across gender, which involves testing whether the measurement model operates equivalently for different gender groups. This analysis allows us to determine if the AT-GLLM and AT-PLLM scales measure the same constructs in the same way for males and females, thereby providing evidence of the scales' structural validity and ensuring that any observed differences are not due to measurement bias. When the criteria for measurement invariance are met, this provides evidence that our constructs have the same structure across groups of males and females (Putnick & Bornstein, 2016).

Invariance testing was conducted sequentially, starting with the least restrictive model of configural invariance, which assesses whether the overall factor structure is consistent across groups. For configural model, a multi-group CFA model is fitted for each group separately, without any equality constraints. This model allows us to test whether the same factorial structure holds across all groups. This was followed by testing for metric invariance to determine whether factor loadings are equivalent across groups, indicating that the construct is measured similarly. The Metric invariance model is a constrained version of the configural model where the factor loadings are assumed to be equal across groups, but the intercepts are allowed to vary between groups.

Subsequently, scalar invariance was examined to evaluate whether item intercepts are equal across groups, enabling valid comparisons of latent means. The Scalar invariance model is a constrained version of the metric model where both the factor loadings and intercepts are assumed to be equal across groups. Finally, strict invariance was tested to assess whether residual variances are invariant, ensuring that measurement errors are consistent across groups. The strict invariance model is a constrained version of the scalar model where the factor loadings, intercepts, and residual variances are fixed across groups (Brown, 2015).

To evaluate whether the null hypothesis of invariance could be maintained, we used the criteria of $|\Delta CFI| \leq .01$ and $|\Delta RMSEA| \leq .015$ (Kim et al., 2017), supported by chi-square difference tests. A change in CFI of less than .01 indicates that the model's fit does not significantly worsen when parameters are constrained to be equal across groups, thereby supporting invariance.

***External validation***

For external validation, we employed the following variables: self-efficacy, acceptance of AI, and fear of AI. Several studies have found that increased self-efficacy correlates with increased acceptance and more consistent use of AI applications (Naiseh et al., 2025; Wang & Chuang, 2024). People who are confident in their technological abilities tend to be more inclined to explore system features, trust the results, and incorporate the technology into their daily activities. Based on these findings, we hypothesize that self-efficacy will significantly predict attitudes toward LLM, with higher self-efficacy leading to more positive perceptions and greater acceptance of LLMs.

The ATAI scale was-included as an external validation tool for two key reasons. First, since ATAI provided the conceptual basis for creating the new LLM attitude scales, validating against it ensures the core construct is maintained. Specifically, we anticipate a stronger association between ATAI and the AT-GLLM scale, given that both measure general attitudes toward AI technologies and are conceptually closely related. We expect a relatively weaker but still meaningful relationship between ATAI and the AT-PLLM scale. Analysing the relationship between ATAI and the LLM-specific scales helps evaluate both convergent and discriminant validity.

To assess the external validity of the AT-GLLM and AT-PLLM scales, we used structural equation modeling (SEM). The SEM model was conducted by specifying four observed variables to serve as dependent outcomes: AT-GLLM acceptance, AT-GLLM fear, AT-PLLM acceptance, and AT-PLLM fear. The analysis involved regressing each outcome on these external predictors: Self-efficacy and two dimensions of ATAI (ATAI acceptance and ATAI fear). All variables were standardized prior to analysis, as they were measured on different scales. The model was estimated via maximum likelihood (ML) using NLMINB as optimization method, implemented through the lavaan R package (version 0.6-19). To obtain robust standard errors and test statistics, nonparametric bootstrapping with 1,000 draws was performed. Model fit was assessed using several fit indices, including CFI, TLI, RMSEA, and SRMR.

# 3. Results

## 3.1. Descriptive statistics and correlations

Table 2 presents the descriptive statistics and Pearson correlations for AT-GLLM and AT-PLLM scales. The descriptive statistics indicated that participants showed higher acceptance and lower fear toward their primary LLM than LLMs in general.

The correlations indicated that higher acceptance of general LLMs was associated with lower fear toward both primary and general LLMs, as well as AI in general. Additionally, AT-GLLM acceptance showed positive and moderate to strong associations with AT-PLLM acceptance, ATAI acceptance, and self-efficacy. Greater fear of general LLMs was strongly linked to higher fear across different scales and was negatively related to acceptance measures. Furthermore, AT-PLLM acceptance was significantly positively related to ATAI acceptance and self-efficacy but negatively related to AT-PLLM fear and ATAI fear. Lastly, fear of specific LLMs was positively associated with fear of AI and negatively associated with ATAI acceptance.

**Table 2.** Descriptive statistics and Pearson correlations between the study variables.

| | Descriptive statistics | | | | Correlations | | | | | |
|---|---|---|---|---|---|---|---|---|---|---|
| Scale | Mean | SD | Skewness | Kurtosis | 1 | 2 | 3 | 4 | 5 | 6 |
| 1. AT-GLLM acceptance | 12.94 | 3.95 | -0.39 | -0.30 | - | | | | | |
| *I trust them.* | 5.90 | 2.49 | -0.48 | -0.26 | | | | | | |
| *They will benefit humankind.* | 7.04 | 2.03 | -0.75 | 0.93 | | | | | | |
| 2. AT-GLLM fear | 12.73 | 6.50 | 0.40 | -0.23 | -.215*** | - | | | | |
| *I fear them.* | 2.68 | 2.58 | 0.85 | -0.09 | | | | | | |
| *They will destroy humankind.* | 3.27 | 2.80 | 0.56 | -0.69 | | | | | | |
| *They will cause job losses.* | 6.78 | 2.58 | -0.71 | -0.09 | | | | | | |
| 3. AT-PLLM acceptance | 13.04 | 3.93 | -0.54 | -0.03 | .781*** | -.141* | - | | | |
| *I trust it.* | 6.08 | 2.39 | -0.67 | -0.04 | | | | | | |
| *It will benefit humankind.* | 6.96 | 2.12 | -0.78 | 0.74 | | | | | | |
| 4. AT-PLLM fear | 11.78 | 5.93 | 0.41 | 0.17 | -.146* | .879*** | -.127* | - | | |
| *I feat it.* | 2.24 | 2.36 | 1.01 | 0.49 | | | | | | |
| *It will destroy humankind.* | 3.01 | 2.66 | 0.63 | -0.50 | | | | | | |
| *It will cause job losses.* | 6.53 | 2.58 | -0.64 | -0.14 | | | | | | |
| 5. ATAI acceptance | 12.82 | 3.69 | -0.36 | -0.24 | .776*** | -.250*** | .733*** | -.223*** | - | |
| *I trust AI.* | 5.83 | 2.24 | -0.40 | -0.13 | | | | | | |
| *AI will benefit humankind.* | 7.00 | 1.98 | -0.53 | 0.10 | | | | | | |
| 6. ATAI fear | 14.88 | 6.42 | 0.13 | -0.68 | -.223*** | .806*** | -.233*** | .748*** | -.264*** | - |
| *I fear AI.* | 3.91 | 2.79 | 0.24 | -0.94 | | | | | | |
| *AI will destroy humankind.* | 3.77 | 2.83 | 0.37 | -0.83 | | | | | | |
| *AI will cause many job losses.* | 7.19 | 2.26 | -0.58 | -0.36 | | | | | | |
| 7. Self-efficacy | 8.04 | 1.54 | -0.54 | -0.40 | .265*** | -.106 | .202** | -.051 | .293*** | -.038 |

## 3.2. Confirmatory factor analysis results

The two-factor models of the AT-GLLM and AT-PLLM were examined using confirmatory factor analyses on the whole sample. The CFA results for both models are reported in Table 3.

**AT-GLLM Scale:** The model resulted in a good fit: $\chi^2(3) = 3.137$, $p = .371$; CFI = 1; TLI = .999; RMSEA = .014; SRMR = .02. All standardized factor loadings were significant ($p < .001$), ranging from .46 to .96, indicating that each observed variable meaningfully contributed to its respective latent construct (see Table 2). The latent factors of acceptance and fear were negatively correlated ($r = -.368$, $p < .001$), suggesting that stronger fear attitudes were moderately associated with lower trust in LLMs. An examination of the model residuals revealed no substantial sources of local misfit, with all residual correlations being small in magnitude. The largest residual correlation was .063, well below the threshold of .10, indicating minimal unexplained covariance and supporting the model's adequacy in capturing the relationships among items while maintaining local independence.

**AT-PLLM Scale:** The results of CFA exhibited a very good fit: $\chi^2(3) = 3.276$, $p = .351$; CFI = .999; TLI = .996; RMSEA = .019; SRMR = .021. All items loaded significantly and meaningfully onto their respective latent factors, with loadings ranging from .38 to .99 (see Table 2). The correlation between the two latent factors was negative ($r = -.261$, $p = .004$), suggesting that lower fear of primary LLMs was associated with higher acceptance. An inspection of the standardized residual covariance matrix revealed no problematic model-data misfit, as all residual correlations were small, with none exceeding ±.10, indicating minimal overlooked dependencies or model misspecification.

**Table 3.** CFA factor loadings and R-squares

| | | AT-GLLM | | AT-PLLM | |
|---|---|---|---|---|---|
| Factor | Item | Loading | $R^2$ | Loading | $R^2$ |
| Acceptance | I trust it/them | .682 | .465 | .517 | .267 |
| | It/They will benefit humankind | .799 | .607 | .992 | .985 |
| Fear | I fear it/them | .720 | .518 | .664 | .441 |
| | It/They will destroy humankind | .952 | .906 | .914 | .836 |
| | It/They will cause job losses | .457 | .209 | .382 | .146 |

**Note:** $R^2$ represents the proportion of variance in each item explained by the factor it loads on.

It has to be noted that in both scales, two items *"They will benefit humankind"* (acceptance) and *"They will cause job losses"* (fear) showed moderate error correlation. This indicates that shared variance between these items not captured by the latent factors and implies an overlap in the way how individuals evaluate these items. However, each of these items showed a strong loading into their factors and the factor structure remains intact. Previous research showed that common latent factors can correlate with variance components not explained by the model without threatening factor validity (Beauducel & Hilger, 2016). Whether the observed coexistence of optimism in benefits of LLMs and worry about losing jobs reflect regional attitudes toward AI as it was reported earlier (Rahman et al., 2024), or it is specific to LLMs, remains an area for further investigations.

## 3.3. Reliability analysis results

Both subscales of AT-GLLM demonstrated acceptable internal consistency, with Cronbach's alpha coefficients of .684 for acceptance (McDonald's omega of .693) and .747 for fear (McDonald's omega of .768). The Average Variance Extracted (AVE) was .522 for the acceptance subscale and .565 for the fear subscale which are above to the recommended threshold. In terms of Composite Reliability (CR), the acceptance dimension scored .687, while the fear dimension scored .771.

Internal consistency was acceptable for both dimensions of AT-PLLM, with Cronbach's alpha values of .680 for acceptance (McDonald's omega of .684) and .677 for fear (McDonald's omega of .707), moderate but suitable for exploratory research. The AVEs were .579 for acceptance and .484 for fear, while the CRs were .717 and .706, respectively. These statistical indicators collectively support the scale's reliability and validity, confirming its consistency and credibility.

## 3.4. Measurement invariance results

Next, we tested measurement invariance to determine whether the factor structures of the AT-GLLM and AT-PLLM scales are equivalent across two groups of males and female.

***AT-GLLM Scale***

The configural invariance model, in which all parameters were estimated freely between two groups, showed excellent fit to the data: $\chi^2(6) = 9.628$, $p = .141$, CFI = .989, TLI = .963, RMSEA = .070, SRMR = .031, indicating that the basic factor structure is consistent across males and females.

Next, we tested the metric invariance model by constraining the factor loadings to be equal across groups which resulted in a very good fit: $\chi^2(9) = 16.593$, $p = .057$, CFI = .977, TLI = .949, RMSEA = .082. Comparing configural and metric models showed that constraining factor loadings did not significantly worsen the model fit ($\Delta\chi^2(3) = 6.965$, $p = .073$; ΔRMSEA = .013; ΔCFI = -.012). This indicates that the scale has equivalent factor loadings across genders. In other words, the constrained model (i.e., metric model) fits the data equally well.

Additionally, when we constrained both the factor loadings and intercepts to be equal, the model still fit well: $\chi^2(12) = 21.463$, $p = .034$, CFI = .971, TLI = .952, RMSEA = .080. Comparing scalar invariance against metric model resulted in slight change in fit indices and non-significant chi-square difference ($\Delta\chi^2(3) = 4.87$, $p = .182$; ΔRMSEA = -.003; ΔCFI = -.006), suggesting scalar invariance is supported.

Lastly, testing the strict invariance model by constraining residual variances to be equal across groups resulted in a modest decrease in fit: $\chi^2(17) = 33.906$, $p = .008$, CFI = .948, TLI = .939,

RMSEA = .090. The decrease in fit indices and the significant chi-square difference ($\Delta\chi^2(3)$ = 12.443, $p$ = .029; ΔRMSEA = .010; ΔCFI = -.023) suggest that residual variances differ between males and females (Table 4).

Overall, the findings support measurement invariance of the AT-GLLM scale at the configural, metric, and scalar levels, permitting meaningful comparison of latent means between males and females, although strict invariance is only partially supported.

***AT-PLLM Scale***

The configural model exhibited a good fit to the data: $\chi^2(6)$ = 5.797, $p$ = .453, CFI = 0.966, TLI = 1, RMSEA = .000, and SRMR = .028. This indicates that the basic factor structure of AT-PLLM scale is appropriate for both groups.

The metric invariance model also showed acceptable fit: $\chi^2(9)$ = 7.485, p = .585, CFI = 0.832, TLI = 1, RMSEA = .037, and SRMR = .037. Comparing the metric invariance model to the configural model showed no significant difference ($\Delta\chi^2(3)$ = 1.687, $p$ = .640; ΔRMSEA = .000; ΔCFI = .000), supporting the assumption of equivalent factor loadings. This finding suggests that after constraining the factor loadings to be equal across groups, the model fit did not change substantially.

When testing for scalar invariance by constraining intercepts alongside factor loadings, the fit remained acceptable: $\chi^2(12)$ = 13.557, $p$ = .332, CFI = .993, TLI = .989, RMSEA = .032, and SRMR = .050. The fit change from the metric model was minimal (ΔCFI = -.007; ΔRMSEA = .032), and the difference in chi-square was non-significant ($\Delta\chi^2$ = 6.073, $p$ = .108), suggesting scalar invariance as well.

Finally, the strict invariance model, which added constraints on residual variances, also fit well: $\chi^2(17)$ = 21.356, $p$ = .175, CFI = .981, TLI = .978, RMSEA = .045, and SRMR = .065. Testing the strict invariance model compared to the scalar model did not significantly worsen the fit ($\Delta\chi^2(3)$ = 7.799, $p$ = .168; ΔRMSEA = .013; ΔCFI = -.012), indicating that residual variances are equivalent across groups (Table 4).

Overall, these results support measurement invariance of the AT-PLLM scale at the configural, metric, scalar, and strict levels, indicating that the AT-PLLM scale functions equivalently across groups and enabling meaningful comparisons.

**Table 4.** Model fit indices and comparisons of measurement invariance models.

| Scale | Model | $\chi^2(df)$ | $\chi^2/df$ | CFI | TLI | RMSEA | SRMR | Comparison | $\Delta\chi^2(p)$ | ΔCFI | ΔRMSEA |
|---|---|---|---|---|---|---|---|---|---|---|---|
| AT-GLLM | M1: Configural | 9.628 (6) | 1.605 | .989 | .963 | .070 | .031 | - | - | - | - |
| | M2: Metric | 16.593 (9) | 1.844 | .977 | .949 | .082 | .044 | M2 vs. M1 | 6.965 (.073) | -.012 | .013 |
| | M3: Scalar | 21.463 (12) | 1.789 | .971 | .952 | .080 | .051 | M3 vs. M2 | 4.870 (.182) | -.006 | -.003 |
| | M4: Strict | 33.906 (17) | 1.994 | .948 | .939 | .090 | .071 | M4 vs. M3 | 12.443 (.029) | -.023 | .010 |
| AT-PLLM | M1: Configural | 5.797 (6) | .966 | 1 | 1.003 | .000 | .028 | - | - | - | - |
| | M2: Metric | 7.485 (9) | .832 | 1 | 1.014 | .000 | .037 | M2 vs. M1 | 1.687 (.640) | 0 | 0 |
| | M3: Scalar | 13.557 (12) | 1.130 | .993 | .989 | .032 | .050 | M3 vs. M2 | 6.073 (.108) | -.007 | .032 |
| | M4: Strict | 21.356 (17) | 1.256 | .981 | .978 | .045 | .065 | M4 vs. M3 | 7.799 (.168) | -.012 | .013 |

**Note:** Δ represents the difference between fit indices.

## 3.5. External validity

To evaluate the external validity of the AT-GLLM and AT-PLLM scales, this analysis investigated how these variables are related to individuals' self-efficacy and their acceptance and fear of AI. SEM was employed to determine whether self-efficacy and general AI attitudes predicted the four LLM-specific attitude factors: acceptance and fear for both GLLM and PLLM. Table 5 shows the SEM results of the AT-GLLM and AT-PLLM scales with the ATAI subscales and participants' self-efficacy.

The overall model showed an excellent fit to the data: $\chi^2(3) = 6.538$, $p = .088$; CFI = .997; TLI = .979; RMSEA = .069; SRMR = .018. A considerable proportion of variance in each outcome variable was captured by the model. It explained 60.5% of the variance in AT-GLLM Acceptance ($R^2 = .605$), 66% in AT-GLLM Fear ($R^2 = .66$), 53.6% in AT-PLLM Acceptance ($R^2 = .536$), and 55.9% in AT-PLLM Fear ($R^2 = .559$). These results suggest that the external predictors (Self-efficacy, ATAI-Acceptance, and ATAI-Fear) jointly offered significant explanatory power for attitudes toward both general and primary LLMs in the Arab sample.

For AT-GLLM acceptance, the SEM analysis showed that confidence in one's abilities (self-efficacy) had a small to none impact on accepting general LLMs ($\beta = .042$, $SE = .048$, 95% $CI$ [-.051, .136], $p = .373$). Individuals' positive attitudes toward AI strongly increased their likelihood of accepting general LLMs ($\beta = .758$, $SE = .043$, 95% $CI$ [.674, .842], $p < .001$). However, ATAI fear was not a significant predictor for AT-GLLM acceptance ($\beta = -.023$, $SE = .044$, 95% $CI$ [-.109, .063], $p = .596$).

For AT-GLLM fear, higher fears about AI were strongly linked to fears about LLMs in general ($\beta = .804$, $SE = .038$, 95% $CI$ [.729, .879], $p < .001$). Lastly, having a positive attitude toward AI ($\beta = -.019$, $SE = .044$, 95% $CI$ [-.105, .068], $p = .673$) and self-efficacy ($\beta = -.070$, $SE = .045$, 95% $CI$ [-.158, .018], $p = .117$) did not significantly influence fears about general LLMs.

Regarding AT-PLLM acceptance, ATAI acceptance remained a significant positive predictor ($\beta = .724$, $SE = .048$, 95% $CI$ [.629, .819], $p < .001$), indicating that people who show positive attitude toward AI are more likely to accept primary LLM. In contrast, confidence in one's abilities ($\beta = -.013$, $SE = .050$, 95% $CI$ [-.110, .085], $p$ = 801) and fears about AI ($\beta = -.042$, $SE = .051$, 95% $CI$ [-.141, .058], $p = .411$) did not significantly affect their acceptance of primary LLM.

Lastly, regarding fear of primary LLMs (AT-PLLM fear), confidence in abilities showed a small, negative, and non-significant affect ($\beta$ = -.017, *SE* = .050, 95% *CI* [-.114, .081], $p$ = .736). Acceptance of AI was not a significant predictor for fear of primary LLMs ($\beta$ = -.022, *SE* = .048, 95% *CI* [-.116, .072], $p$ = .653). However, fears about AI were strong and significant predictor of fears about primary LLMs ($\beta$ = .742, *SE* = .043, 95% *CI* [.658, .825], $p$ < .001), indicating that people with higher fears about AI tend to also have greater fears of primary LLM.

**Table 5.** Standardized estimates of regressing AT-GLLM and AT-PLLM scales against ATAI and self-efficacy

| | | | | | | 95% *CI* | | |
|---|---|---|---|---|---|---|---|---|
| Dependent | Independent | Estimate | Std. error | *z*-value | *p*-value | Lower | Upper | $R^2$ |
| AT-GLLM acceptance | Self-efficacy | .042 | .048 | .891 | .373 | -.051 | .136 | .605 |
| | ATAI acceptance | .758 | .043 | 17.681 | < .001*** | .674 | .842 | |
| | ATAI fear | -.023 | .044 | -.530 | .596 | -.109 | .063 | |
| AT-GLLM fear | Self-efficacy | -.070 | .045 | -1.596 | .117 | -.158 | .018 | .660 |
| | ATAI acceptance | -.019 | .044 | -.422 | .673 | -.105 | .068 | |
| | ATAI fear | .804 | .038 | 21.006 | < .001*** | .729 | .879 | |
| AT-PLLM acceptance | Self-efficacy | -.013 | .050 | -.252 | .801 | -.110 | .085 | .536 |
| | ATAI acceptance | .724 | .048 | 15.003 | < .001*** | .629 | .819 | |
| | ATAI fear | -.042 | .051 | -.822 | .411 | -.141 | .058 | |
| AT-PLLM fear | Self-efficacy | -.017 | .050 | -.337 | .736 | -.114 | .081 | .559 |
| | ATAI acceptance | -.022 | .048 | -.450 | .653 | -.116 | .072 | |
| | ATAI fear | .742 | .043 | 17.357 | < .001*** | .658 | .825 | |

**Note:** $R^2$ represents the proportion of variance in each dependent variable accounted for by all predictors in the SEM.

We also examined residual covariances among the four LLM attitude subscales to assess unexplained overlap after controlling for ATAI subscales and self-efficacy. A moderate residual covariance was found between AT-GLLM Acceptance and AT-PLLM Acceptance ($\beta_{resid}$ = .493, $p$ < .001), suggesting that general and primary LLM acceptance share common variance not captured by the predictors. A stronger residual covariance emerged between AT-GLLM Fear and AT-PLLM Fear ($\beta_{resid}$ = .698, $p$ < .001), indicating substantial unexplained overlap in fear responses across both LLM contexts. All other residual covariances were non-significant ($p$ > .05), suggesting minimal unexplained association beyond the modelled predictors.

# 4. Discussion

With growing awareness of LLMs, it is crucial to develop reliable, culturally appropriate instruments to accurately measure these attitudes in different populations. This study aimed to address this need by validating two scales to measure attitude to LLM in the Arab population - the Attitudes Toward General Large Language Models (AT-GLLM) and the Attitudes Toward Primary Large Language Models (AT-PLLM) originally developed in the UK population (Liebherr et al., 2025b).

In line with the original UK validation by Liebherr et al. (2025b), our study demonstrated that both the AT-GLLM and AT-PLLM scales retained a two-factor structure, comprising Acceptance (2 items) and Fear (3 items), in the Arab sample. Importantly, this factor configuration aligns with theoretical frameworks of attitudes toward AI, which distinguish positive (acceptance/trust) and negative (fear/concerns) (Sindermann et al., 2021). Moreover, the measurement invariance testing demonstrated that the AT-GLLM and AT-PLLM scales function equivalently for males and females in the Arab sample, which is essential for valid cross-gender comparisons of latent attitudes. This indicates that respondents interpret and respond to items similarly across gender, and for participants with the same underlying level of Acceptance or Fear, their average responses to the items do not differ simply due to gender.

However, when examining whether the residual 'noise' (i.e., the part of each item's response not explained by the latent traits) was consistent across gender, we found that AT-PLLM performed equally well for males and females, demonstrating equal measurement precision. In contrast, AT-GLLM showed differences in residual variance between genders, suggesting that item responses for attitudes to general LLM include a larger variability in one gender. Though this does not undermine comparisons of the latent constructs, it does indicate that in the Arab sample males and females tend to respond more consistently about attitudes toward primary LLM compared to attitudes toward general LLMs. The original scales (Liebherr et al., 2025b) developed using a sample demographically similar to ours with respect to age and gender found no gender differences at the item level for attitudes toward general LLMs. The fact that we did observe such differences suggests the possibility of cultural variation in item responses. However, to date, there is limited empirical research directly comparing attitudes toward primary versus general LLMs or across different cultural contexts. Therefore, we suggest focusing on latent constructs (i.e., LLM Acceptance and LLM Fear) rather than individual item scores when using the AT-GLLM scale, as latent measures provide more reliable.

Compared to the ATAI framework (i.e., the theoretical and methodological foundation for developing original AT-PLLM and AT-GLLM) across German, Chinese, and UK samples, the internal reliability of our Arab sample is well within the expected range. The ATAI work reported Cronbach's alpha values for acceptance between .64 and .73 and for fear between .61 and .66, acknowledging that these are acceptable given the small number of items per scale (Sindermann et al., 2021). In the Arab sample, acceptance ranged from .680 to .684 and fear from .677 to .747, demonstrating that our results are in line with, and in some cases slightly exceed, the reliabilities observed in the foundation work.

When set against the original UK validation of the AT-GLLM and AT-PLLM scales, our Arab sample showed slightly lower reliability overall. For AT-GLLM, alpha values in the Arab sample were .684 for acceptance and .747 for fear, compared with .786 and .776 in the UK. Composite reliability and Average Variance Extracted values in our data were still within acceptable ranges, though slightly lower than the UK figures, particularly for acceptance. For AT-PLLM, alpha values in the Arab sample (.680 acceptance, .677 fear) fell below the UK's (.749 and .742), and CR values were modestly reduced but remained at levels supporting construct reliability. Notably, Average Variance Extracted for acceptance was higher in our Arab sample (.579 vs .470 in the UK), suggesting stronger shared variance for that dimension, while the fear subscale AVE (.484 vs .620) still indicated an adequate relationship between items and their underlying construct. It is worth noting that small differences in internal consistency between the Arab and UK samples are not unexpected. The UK validation was based on a substantially larger sample (n = 526), which naturally produces more stable reliability estimates (Bonett, 2002b). In contrast, slightly lower or more variable alpha values in smaller samples are a normal statistical occurrence and do not indicate problems with measurements (Iacobucci & Duhachek, 2003; Bujang et al., 2018).

At the item level, our validation in the Arab sample demonstrates that the overall factor loading pattern of the AT-PLLM and AT-GLLM scales (Liebherr et al., 2025b) is preserved, with some notable shifts that reflect the cultural context. For AT-GLLM, "I trust it/them" loaded at 0.682 in the Arab sample compared to 0.868 in the UK, indicating that trust plays a less central role in defining acceptance in our data. In contrast, "benefit humankind" showed only a modest difference (0.799 vs 0.734), supporting the stability and consistency of this belief across both cultures. Within the fear dimension, "I fear it/them" was lower in our sample (0.720 vs 0.915), yet "destroy humankind" was markedly higher (0.952 vs 0.774), highlighting a sharper expression of existential fears among Arab respondents. Notably, the ATAI framework reported even lower loadings for this item (i.e., "destroy humankind") across Germany, UK, and China (0.64-0.70) (Sindermann et al., 2021). The "job losses" item remained relatively weak in both our sample (0.457) and the original scale (0.463) (Liebherr et al., 2025b), though still substantially higher than in the ATAI framework (0.23-0.42) (Sindermann et al., 2021). In summary, our validation confirms that the Arab sample retains the same general factor structure as the UK validation and the ATAI framework, while revealing meaningful shifts: acceptance is slightly less defined by trust but more strongly by perceived societal benefits, and fear is anchored more heavily in existential concerns. We summarised these findings in Table 5, providing recommendations to guide the practical use of the AT-GLLM scale in the Arab population.

Our validation results of the AT-PLLM in the Arab sample shows the same 2-factor structure as the original scale (Liebherr et al., 2025b), but with several differences in item contributions to each factor. Within the acceptance factor, "I trust it/them" loaded at 0.517 in the Arab sample compared to 0.758 in the UK, indicating that trust plays a less central role in defining acceptance of the primary LLM in the Arab population. In contrast, "benefit humankind" was considerably stronger in the Arab sample (0.992) than in the UK (0.791), suggesting that perceptions of societal benefit are a particularly powerful driver of acceptance for the primary LLM in this context. In the fear dimension, "I fear it/them" loaded at 0.664 in the Arab sample versus 0.819 in the UK, indicating a moderately weaker association with fear in Arab data. However, "destroy humankind" was almost identical between the two samples (0.914 vs 0.912), confirming its robustness as a core fear indicator across contexts. The "job losses" item remained weak in both settings (0.382 in the Arab sample, 0.439 in the UK), suggesting its more peripheral role in defining fear in the attitude to user' primary LLM. Table 5 provided recommendations to guide the practical use of the AT-PLLM scale in the Arab population.

**Table 5.** Item-level guidance for practitioners for applying the AT-GLLM and AT-PLLM scales in the Arab population

| Item | Item loading | | Interpretation and recommendation for practical use |
|---|---|---|---|
| | Arab | UK | |
| **Attitudes toward general LLM (AT-GLMM)** | | | |
| I trust it/them | 0.682 | 0.868 | Trust is a weaker marker of acceptance in the Arab population. When using the scale in Arab populations, focus on supplementing trust-related items with other acceptance indicators, as trust alone may not fully capture positive attitudes toward AI. |
| It/They will benefit humankind | 0.799 | 0.734 | Perceived societal benefits are a stable core of acceptance across Arab and UK samples. In practice, messages and interventions should emphasise these benefits to facilitate acceptance. |
| I fear it/them | 0.72 | 0.915 | Fear is a somewhat weaker predictor in the Arab sample compared to the UK. Practitioners should recognise that personal fear may be present but not as dominant as existential concerns. |
| It/They will destroy humankind | 0.952 | 0.774 | Exceptionally strong in the Arab population, suggesting existential fears are a central component of negative attitudes toward AI. Addressing and mitigating such fears should be a priority in communication and policy work. |
| It/They will cause job losses | 0.457 | 0.463 | Consistently weak in both contexts, indicating that job-loss concerns are less central to fear of AI. This item should be retained for comparability but not relied on as a primary fear indicator in Arab population. |
| **Attitudes toward primary LLM (AT-PLMM)** | | | |
| I trust it/them | 0.517 | 0.758 | Trust is a weaker marker of acceptance for the primary LLM in the Arab population than in the UK. Practitioners should combine trust-related measures with other acceptance indicators when assessing attitudes. |
| It/They will benefit humankind | 0.992 | 0.791 | Perceived societal benefits are a particularly strong driver of acceptance in the Arab population. Messaging and policy interventions should emphasise these benefits to maximise acceptance. |
| I fear it/them | 0.664 | 0.819 | Fear is somewhat less tied to personal apprehension toward primary LLM in the Arab population than in the UK, though it remains relevant. Practitioners should address fear but recognise that existential concerns are more central. |
| It/They will destroy humankind | 0.914 | 0.912 | Consistently strong in both populations, confirming its role as a core fear indicator. Interventions aimed at reducing fear should directly address these existential concerns in the Arab population. |
| It/They will cause job losses | 0.382 | 0.439 | Weak in both populations, indicating that job-loss concerns are less central to fear. This item should be retained for comparability but is not a primary driver of fear in the Arab population. |

In both the AT-PLLM and AT-GLLM scales, modification indices indicated an error correlation between the items “It/They will benefit humankind” and “It/They will cause job losses.” This suggests that responses to these two items share variance beyond what is explained by the acceptance and fear factors alone. For practitioners working with the Arab population, this means that participants may be linking perceptions of societal benefit with concerns about employment impact, possibly seeing them as part of the same mental evaluation of LMM’s broader consequences. While this correlation does not undermine the validity of the scales, in applied use, shifts in one perception (e.g., perceived benefits) may be accompanied by shifts in the other (e.g., job loss concerns). Researchers and practitioners should be aware of this interplay when interpreting scores, particularly in policy, communication, or interventions.

Our external validation analysis demonstrated that both AT-GLLM and AT-PLLM scales function coherently in the Arab sample, aligning with broader AI attitudes. In this population, ATAI Acceptance was the sole significant predictor of both general and primary LLM acceptance, while ATAI Fear was the sole predictor of fear toward both scales. Self-efficacy did not emerge as a significant predictor in any of the four model’ paths once ATAI dimensions were accounted for. However, acceptance of general AI accounted for only 60.5 % of the variance in AT-GLLM Acceptance, and fear of general AI accounted for 66 % of fear toward general LLMs; the proportions were even lower for the primary LLM, with 53.6 % of acceptance and 55.9 % of fear explained. Thus, between 34 % and 46 % of the variance remains specific to attitudes toward LLM even after controlling for general AI attitudes. By comparison, the UK sample models explained more variance ranging between 62.3 and 70.6 % across attitudes toward general and primary LLM. Notably, self-efficacy had a small but significant positive association with AT-PLLM Acceptance in the UK, whereas in the Arab sample self-efficacy did not contribute significantly. Recent empirical studies in Arab populations have shown that self-efficacy plays a key role in shaping attitudes toward AI (Naiseh et al., 2025). This is consistent with our own findings, where self-efficacy had a stronger correlation with ATAI Acceptance than with attitudes toward LLMs.

The validated AT-GLLM and AT-PLLM scales could be applied in several ongoing AI initiatives in the Arab region to help understand public attitudes toward large language models. For example, the UAE’s TII-ATRC Falcon Arabic project, which develops Arabic-focused LLMs, could use these tools to assess levels of acceptance and concern among different groups. Similarly, Arab institutes and programs, such as Saudi Arabia’s Humain initiative (Arab News, 2025) which aims to create multimodal Arabic LLMs, could employ the scales to monitor changes in public attitudes as the technology develops by specifying the specific AI system of interest in the instruction to participants. Beyond government-led projects, these measures could be valuable in education to track student and faculty responses to LLM integration in teaching and research. In healthcare, the scales could be used to gauge acceptance and fears surrounding the adoption of LLM-based tools for patient communication, clinical documentation, and decision support in Arabic-speaking contexts.

The AT-GLLM and AT-PLLM scales can be employed in surveys, research projects, and longitudinal studies. We recommend incorporating these scales into broader assessment frameworks that will provide a more comprehensive understanding of how LLM attitudes relate to other psychological and behavioural measurements.

# 5. Conclusion

The present study demonstrates that the AT-GLLM and AT-PLLM are valid and reliable measures for assessing acceptance and fear of large language models in Arab populations. Their stable factor structure, satisfactory reliability, and coherent relationships with broader AI attitudes indicate that they capture meaningful and interpretable constructs. These tools can be applied in research and practice to monitor public attitudes toward general LLMs and user's primary LLM, inform policy and communication strategies, and guide interventions.

**Acknowledgement**

Open Access funding provided by the Qatar National Library. This publication was supported by NPRP 14 Cluster grant # NPRP 14C-0916–210015 from the Qatar National Research Fund (a member of Qatar Foundation). The findings herein reflect the work and are solely the responsibility of the authors.

**Authors' Contribution**

BB: Designed, performed and reported the analysis and wrote the first draft.
AY: Conceptualized and designed the study, mentored and validated the analysis, contributed to the first draft, and reviewed and edited the paper.
SA: Translated the scale, curated the data, validated the analysis, reviewed and edited the paper.
CSMH: Validated the analysis, reviewed and edited the paper.
GX: Validated the analysis, reviewed and edited the paper.
RA: Conceptualized and designed the study, translated the scale, validated the analysis, and reviewed and edited the paper.

**Ethics Declarations**

***Ethics approval***

This study was approved by the Ethics Committee at Bournemouth University, UK (N62239, 03.03.2025).

***Consent to Participate***

All participants were provided with detailed information about the study prior to participation. Informed consent was obtained from all individuals before their inclusion in the study. They were also made aware of their right to withdraw from the study at any time. They consented for the anonymous responses to be made publicly available. Participants who completed the survey received monetary compensation for their time

### *Consent to Publication*

All participants consented to the use of anonymized data for research dissemination, including publication in scientific journals.

### *Competing Interest*

The authors declare no conflict of interest.

### *Data Availability*

Supplementary materials and dataset are available on the Open Science Framework (OSF) at the following link: https://osf.io/p7sg2/

# Appendix 1

## The Attitude towards General LLM (AT_GLLM) Scale

*Think of LLMs in general and the applications built on them. To what extent does the following statement apply to you?*

| Item nr | | 0 (Strongly Disagree) | 1 | 2 | 3 | 4 | **5** | 6 | 7 | 8 | 9 | 10 (Strongly Agree) |
|---|---|---|---|---|---|---|---|---|---|---|---|---|
| 1 | I fear them | | | | | | | | | | | |
| 2 | I trust them | | | | | | | | | | | |
| 3 | They will destroy humankind | | | | | | | | | | | |
| 4 | They will benefit humankind | | | | | | | | | | | |
| 5 | **They will cause many job losses** | | | | | | | | | | | |

**Scoring:**

The AT_GLLM measures two components: Acceptance of LLMs and Fear of LLMs.

Items are rated on an 11 -point Likert scale ranging from 0 (Strongly Disagree) to 10 (Strongly Agree). To support user experience and reduce cognitive load, midpoint 5 is visually emphasized in bold.

Acceptance score is calculated by summing up Item 2 and Item 4 and Fear score is calculated as the sum of responses to items 1, 3 and 5. Higher scores indicate stronger endorsement of the respective attitude, whether Acceptance or Fear.

## النسخة العربية من مقياس الموقف تجاه نماذج اللُّغة الكبيرة بشكل عام (AT_GLLM)

فكّر في نماذج اللُّغة الكبيرة (LLM) **بشكل عام** والتطبيقات المبنية عليها. ما مدى انطباق العبارات التالية عليك؟

| **رقم البند** | | 0 (أعارض بشدة) | 1 | 2 | 3 | 4 | **5** | 6 | 7 | 8 | 9 | 10 (أوافق بشدة) |
|---|---|---|---|---|---|---|---|---|---|---|---|---|
| 1 | أشعر بالخوف منها | | | | | | | | | | | |
| 2 | أثق بها | | | | | | | | | | | |
| 3 | ستدمّر البشرية | | | | | | | | | | | |
| 4 | ستفيد البشرية | | | | | | | | | | | |
| 5 | ستتسبب في فقدان العديد من الوظائف | | | | | | | | | | | |

***التقييم:***

AT_GLLM يقيس عنصرين: قبول نماذج اللُّغة الكبيرة - الخوف من نماذج اللُّغة الكبيرة

يتم تقييم البنود على مقياس ليكرت من 11 نقطة يتراوح من 0 (أعارض بشدة) إلى 10 (أوافق بشدة). لدعم تجربة المستخدم وتقليل الإجهاد الذهني، يتم تمييز النقطة المتوسطة 5 بصريًا بخط عريض.

يتم حساب درجة القبول من خلال جمع استجابات البند 2 والبند 4، بينما يتم حساب درجة الخوف بجمع استجابات البنود 1 و 3 و 5. تشير الدرجات الأعلى إلى تأييد أقوى للموقف المعني، سواء كان القبول أو الخوف.

# Attitude towards Primary LLM (AT_PLLM) Scale

*Think of your primary LLM, meaning the LLM you use the most. To what extent does the following statement apply to you?*

| Item nr | | 0 (Strongly Disagree) | 1 | 2 | 3 | 4 | **5** | 6 | 7 | 8 | 9 | 10 (strongly agree) |
|---|---|---|---|---|---|---|---|---|---|---|---|---|
| 1 | I fear it | | | | | | | | | | | |
| 2 | I trust it | | | | | | | | | | | |
| 3 | It will destroy humankind | | | | | | | | | | | |
| 4 | It will benefit humankind | | | | | | | | | | | |
| 5 | It will cause many job losses | | | | | | | | | | | |

**Scoring:**

The AT_PLLM measures two components: Acceptance of Primary LLM and Fear of Primary LLM.

Items are rated on an 11-point Likert scale ranging from 0 (Strongly Disagree) to 10 (Strongly Agree). To support user experience and reduce cognitive load, midpoint 5 is visually emphasized in bold.

Acceptance score is calculated by summing up Item 2 and Item 4 and Fear score is calculated as the sum of responses to items 1, 3 and 5. Higher scores indicate stronger endorsement of the respective attitude, whether Acceptance or Fear.

# النسخة العربية من مقياس الموقف تجاه نموذج اللُّغة الكبير أساسيّ الاستخدام (AT_PLLM)

فكّر في نموذج اللُّغة الكبير (LLM) الذي تستخدمه في الغالب وبشكل أساسيّ. ما مدى انطباق العبارات التالية عليك؟

| رقم البند | | أعارض ) 0 بشدة) | 1 | 2 | 3 | 4 | **5** | 6 | 7 | 8 | 9 | أوافق ) 10 بشدة) |
|---|---|---|---|---|---|---|---|---|---|---|---|---|
| 1 | أشعر بالخوف منه | | | | | | | | | | | |
| 2 | أثق به | | | | | | | | | | | |
| 3 | سيدمّر البشرية | | | | | | | | | | | |
| 4 | سيفيد البشرية | | | | | | | | | | | |
| 5 | سيتسبب في فقدان العديد من الوظائف | | | | | | | | | | | |

***التقييم:***

AT_PLLM يقيس عنصرين: قبول نموذج اللُّغة الكبير أساسيّ الاستخدام و الخوف من نموذج اللُّغة الكبير أساسيّ الاستخدام

يتم تقييم البنود على مقياس ليكرت من 11 نقطة يتراوح من 0 (أعارض بشدة) إلى 10 (أوافق بشدة). لدعم تجربة المستخدم وتقليل الإجهاد الذهني، يتم تمييز النقطة المتوسطة 5 بصريًا بخط عريض.

يتم حساب درجة القبول من خلال جمع استجابات البند 2 والبند 4، بينما يتم حساب درجة الخوف بجمع استجابات البنود 1 و 3 و 5. تشير الدرجات الأعلى إلى تأييد أقوى للموقف المعني، سواء كان القبول أو الخوف